# LANGUAGE EVOLUTION AND POPULATION DYNAMICS IN A SYSTEM OF TWO INTERACTING SPECIES

Kosmas Kosmidis[a] , John M. Halley[b], Panos Argyrakis[a]

[a] Physics Department, University of Thessaloniki, 54124 Thessaloniki, Greece

[b] Department of Biology, University of Thessaloniki, 54124 Thessaloniki, Greece

Corresponding Author:

Panos Argyrakis

Physics Department

University of Thessaloniki,

54124 Thessaloniki,

Greece

Tel: 0030-2310-998043

Fax: 0030-2310-998042

e-mail: panos@physics.auth.gr

## ABSTRACT

We use Monte Carlo simulations and assumptions from evolutionary game theory in order to study the evolution of words and the population dynamics of a system comprising two interacting species which initially speak two different languages. The species are characterized by their identity, vocabulary and have different initial fitness, i.e. reproduction capability. The questions we want to answer are: *a.* Will the different initial fitness lead to a permanent advantage? *b.* Will this advantage affect the vocabulary of the species or the population dynamics? *c.* How will the spatial distributions of the species be affected? Does the system exhibit pattern formation or segregation? We show that an initial fitness advantage, although is very quickly balanced, leads to better spatial arrangement and enhances survival probabilities of the species. In most cases the system will arrive at a final state where both languages coexist. However, in cases where one species greatly outnumbers the other in population and fitness, then only one species survives with its "final" language having a slightly richer vocabulary than its initial language. Thus, our results offer an explanation for the existence and origin of *synonyms* in all currently spoken languages.

## INTRODUCTION

Language is a human trait of considerable importance. Human language is not static. Understanding the way it changes and the way it affects the biological evolution of the human species is a very interesting task. The evolution of language is a subject of interdisciplinary interest and has attracted the attention of philosophers, linguists [1, 2], physicists [3] and biologists [4]. Recently, mathematical and computational modeling [5, 6], has been applied to the language evolution of a system (i.e. a group of people) with one common language. This allowed the derivation, in very simple terms, of some important aspects of language. For example, transmission errors limit the number of words that can be reliably communicated and hence the lexicon size of an oral language, but the organization of the language into higher levels of syntax allows this limit to be transcended. This has a formal equivalence with Shannon's information theory, where language becomes a kind of 'noisy channel' [7]. Even for indefinitely large populations not subject to transmission errors, the combined lexicon size cannot be infinite. Given finite learning-time and capacity, population-dynamic arguments demonstrate that only a finite set of words can survive [8]. However, the models that are used to describe this evolution are based on a homogenous distribution. Thus, the spatial arrangement of the system is not considered. In most cases, furthermore, the number of people comprising the system is also kept constant. Very recently, a particularly interesting attempt was made to study the competition between languages [9]. An extension of this simplified model considering spatial effects by means of reaction diffusion equations is already presented in [10]. In spite of their interesting results, both models depend on the assumption that from two competing languages an individual will adopt only one, based on the current "status" of the language [10]. Also, an individual may change from one language to the other and forget his previous choice. Language adoption in reality is, however, a complex task done gradually by word learning. Moreover, the above models do not allow the possibility of an individual being bilingual, and they do not consider population dynamics since the number of individuals is constrained to be constant.

In this paper we use Monte Carlo Simulations to study the language evolution and the population dynamics of a system comprising two “species” *A* and *B*, which are initially randomly distributed in space. Initially the species speak two different languages. The first species speaks language *A*, which for the present study simply means that all such individuals initially have a common vocabulary of 10 words. The second species initially speak language *B*. These also have a common vocabulary of 10 words but these words are different from the words of language *A*. There is no overlap between the words of the two languages. The species are allowed to move on the lattice performing random walks. They also interact with their neighbors as they are both able to learn words that are unknown to them or forget words that are in their vocabulary. Note that the vocabulary does not characterize the species identity. An *A*-individual for example may learn or forget words from his/her language, resulting in speaking a different language, but his/her identity remains the same. He/She always belongs to the *A*-species.

The individuals are also able to reproduce themselves leaving offspring at their neighborhood and they die at a constant rate. Interaction between species means communication. Following the basic assumption of the evolutionary game theory, we assume that correct communication conveys biological fitness. Individuals that communicate more successfully have increased survival probabilities as they leave more offspring. An alternative interpretation is that correct communication will increase the "social" status of the individual. Thus, he will leave more “followers” that will adopt his terminology and his ideas. In a broader sense the same formalism can be used to study the propagation of ideas as well. Monte Carlo simulations based on the above simplified model reveal a surprisingly rich behavior which is difficult to predict *a priori*. There are three main groups of questions that arise while studying such a system of two species.

First, there are questions concerning population dynamics: *a.* Will the different initial fitness lead to a permanent advantage? *b*. Is there a difference in considering a system of two species with the same population and different initial fitness and in considering a system of species with the different initial population and the same initial fitness? Or the two problems are connected and by finding the solution of the first we also solve the second.

Then, there are questions concerning the language evolution: *a.* How does the vocabulary of the species evolve? Will an initial fitness advantage affect the vocabulary of the species and how? How many words survive from each language in an evolutionary stable system?

And finally there are questions concerning the spatial arrangement of the species: *a.* how will the spatial distributions of the species be affected? *b.* Does the system exhibit pattern formation or segregation? *c.* If yes how can we quantify it?

The present study is motivated by a real world situation. Namely, to estimate the impact on the vocabulary of a society that accepts a significant number of emigrants.

## METHODS

We build a model to study the time evolution of a system of two species with different languages via Monte Carlo simulations. The space considered is a discrete lattice in two dimensions, even though our arguments can be independent of the particular space used. The size of the system is 100x100 units (sites). Each site of the lattice can either be empty or it can be occupied with “individuals” *A* or with “individuals” *B*. Initially, each site of the grid is either empty or occupied by species *A* with a probability $c_A$ or by species *B* with a probability $c_B$. Thus, each individual, *i*, is characterized by its species identity, by its fitness $f_i$, which is a measure of the individual relative capability for reproduction, and by its vocabulary. The vocabulary *V* of an individual consists of an array of 20 elements. An element has a value of 1 if the corresponding word is known to the individual or 0 if the word is unknown. Initially, each *A* individual has a vocabulary that is

$$V_A(i) = \begin{Bmatrix} 1, i \leq 10 \\ 0, i > 10 \end{Bmatrix} \quad (1)$$

and each B individual has initially a vocabulary

$$V_B(i) = \begin{Bmatrix} 0, i \leq 10 \\ 1, i > 10 \end{Bmatrix} \quad (2)$$

Individuals belonging to each species are subject to the following activities: Movement, communication, reproduction and mortality.

1) *Movement*. An individual *i* is randomly selected, by use of random numbers. One of the four nearest neighbor sites of this individual is also randomly selected. In case the neighboring site is empty then the individual moves to this site. If it is occupied by another individual *j* then the two communicate (interact) as described below.

2) *Communication by language*. In the event that the destination site is occupied, communication occurs. This communication confers fitness to both individuals according to the number of words individual-*i* has in common with the individual-*j* in the neighboring cell. Specifically:

a) The payoff for the interaction is equal to the number of words *i* and *j* have in common. (e.g., three common words means a payoff of *3*). This payoff value is added to the fitness of each individual, as a reward for successful communication.

b) A *learn-forget* process follows. If individual *i* does not know a word which is known to *j* then there is a probability *p* that *i* will learn it from *j*. If this specific word is learned, then the corresponding array element of his vocabulary will turn from *0* to *1*. However, there is also a probability *q* that *j* will forget this word not known to *i*. Similarly, for words which are known to *i* and unknown to *j*. Thus, words that are unknown to the majority of the population have increased probability of being lost from the language.

3) *Reproduction.* At any instance after the motion of an individual, there is a probability $p_r$ that a reproduction event will take place. Thus, if a new individual is to be born, the selection of the individual to be reproduced is not random but proportional to the individual fitness. This means that individuals with large fitness have a higher probability for reproduction.

We use two different models of selection according to fitness. The first model is to select individual *k* from the lattice with probability $p_k = \frac{f_k}{\sum_k f_k}$, where $f_k$ is the fitness of individual *k* and the sum is over all individuals. The fact that we normalize over the total fitness implies that there is information available to all individuals about the fitness status of their society. For this reason this model is referred as the "global" model. In the second model information about the individual fitness is limited to the Moore Neighborhood (eight nearest sites) as follows: In case a reproduction event will take place one site of the lattice is selected at random. We check the site and its Moore neighborhood to verify that it contains at least one individual. If no individuals reside in any of the nine sites then a new random choice of a site is made, until there is at least one individual in this set of nine sites. From these nine sites an individual is selected for reproduction with probability $p_k = \frac{f_k}{\sum_k f_k}$, where $f_k$ is the fitness of the individual and the sum is over the fitness of the individuals located in the area comprised by the selected site and its Moore Neighborhood. We refer to this second model as the "local" model. The main conceptual difference between these two models is that in the "local" model an individual regulates his genetic activity based on information about his "social" status that comes only from his immediate neighborhood, while in the "global" model this information comes from the complete "society". In both models individuals with higher value of fitness have increased probability of being selected for reproduction. A single offspring is created and placed at a random in the Moore Neighborhood of the parent, if and only if this site is empty. If the selected site is occupied then the reproduction procedure is aborted. The offspring inherits his father’s identity (*A* or *B*), all his vocabulary and a certain percentage, *r*, of his fitness. Also, the father’s fitness is decreased to the same factor *r* of its previous value. Thus, there is a cost to reproduction.

4) *Mortality*. At any instance, after checking for the reproduction process there is a probability $p_d$ that an individual will die, and if this happens he/she is removed from the lattice. The selection of the individual to die is completely random.

5) After each selection for movement of an individual $i$ the time is incremented by $1/n_s$, where $n_s$ is the total number of individuals in the lattice. Thus, one time unit or Monte Carlo Step (MCS) statistically represents the time necessary for each individual to move once. The simulation continues until a prescribed total time is reached.

For statistical purposes we average our results over a large number of realizations. In most cases the time evolution of the system is followed up to 6000 MCS. In all simulation results presented in the present manuscript we have used the values $p$=0.1, $q$=0.1, $p_r$=0.1, $p_d$=0.05, $r$=0.80

Using the above parameters, the average lifetime of an individual in this system turns out to be equal to 19 MCS. (It is the mean of a negative binomial distribution with $n$=1 and $p_d$=0.05). Thus, our study is limited to a time period of approximately 320 generations, unless mentioned otherwise.

## RESULTS AND DISCUSSION

### a. Homogeneous system with constant number of individuals

Initially we consider a model where the birth and death rates are always zero. In this case the population is kept constant. This system is similar to the one studied by Patriarca et al. in [10]. We start with a system comprised of two species with the same concentration and the same initial fitness, randomly distributed on a square lattice. The species move as described in the previous section. We monitor the average number of words that individuals of both species know, as a function of time. The species are allowed to gain fitness through successful communication. We monitor the average fitness per individual, $f$, as a function of time.

The language evolution for this case is presented in Figure *1a*. We calculate the average number of words of each language that individuals of each species know and plot it as a function of time for a system with concentration $c=0.15$, for both A and B species. Thus, we sum the number of words of Language *A* that the *A* individuals know at time t and we divide it by the number of *A*-individuals. This is the "Species *A* -Language *A*" curve. We repeat the same procedure for 150 realizations and take the statistical average. From Figure *1* we see that the species start with a vocabulary of 10 words of their own language and they end up with a vocabulary of (on average) 10 words, five of their own language and five of the other species language. Thus, the population becomes homogeneous, its vocabulary and the number of words is maintained. This is anticipated, since a rich vocabulary is not in this case a competitive evolutionary advantage. The average fitness $f$ is presented in Figure 1b. Note that the curves are almost straight lines in a linear plot (figure not shown) giving the impression that the fitness increase is linear but this is not correct. This is shown if we plot the same data using logarithmic axes as in Figure *1b.* We may understand the form and slope of the curves as follows: The change $df$ in time $dt$ depends on the probability that the individual will meet another individual and on the payoff received by their interaction. Suppose that an individual moves for a time interval $dt$. The probability per unit time, $C$, that he/she meets another individual is proportional to the number of individuals, provided that the distribution of individuals is random. The average payoff gained is not constant in time because there is word learning. We denote it with $p(t)$. Since the number of "collisions" in time $dt$ is expected to be $Cdt$, the fitness increase d$f$ will be

$$\frac{d\,f}{dt} = C\,p(t) \qquad (3)$$

where $C$ is expected to be constant, since the number of individuals is constant here and $p(t)$ is expected to increase with time as the learn-forget process described in the algorithm favors the preservation of common words.

In order to understand why $p(t)$ is increasing with time consider the following: Since in the simulation process one Monte Carlo step corresponds to the time where each individual moves once and since the number of words in the vocabulary of the species was set equal to 10, we expect that $p(t)=10$ for small t. At the very initial stage of the system, one individual might encounter an individual of the same species so that the total fitness of the system increases by 20 points (10+10), or an individual may encounter one of the opposite species so that the total fitness does not increase. The average fitness change per transaction is thus 10 points. At long times the species develop a common vocabulary of 10 words. Then, the average fitness increase per transaction is 20 points. Thus, we expect a gradual increase of the payoff with time and this is verified by the simulation.

**b. Homogenous System with variable number of individuals**

Next we consider the case of two species having exactly the same initial fitness but we allow reproduction and death processes as described in the Methods section. We expect that the language evolution of the system to reach a steady state when all individuals obtain a common vocabulary, as in that case the "learn – forget" process mentioned above stops. For all the simulations of a homogeneous system presented in the present paper we have used a 100x100 square lattice with initial concentration $c_A=c_B=0.15$ and initial fitness $f_A=f_B=30$. There are three important processes to consider, each with its own time-scale.

(i) S*tabilization of the total population number*. Figure *2a* and *b* show plots of the number of individuals and fitness per individual vs. time for both species according to the “local” model. The population dynamics are characterized by a rapid initial increase followed by a steady state at a value of approximately 2600. This is less than the full capacity of the lattice (5000 for each individual). The initial increase of the population is due to the fact that the lattice is relatively empty and there is room for the individuals to place their offspring. The effective birth rate exceeds the death rate and there is a population increase until the point where the lattice becomes rather overcrowded. At this point there is no room for the individuals to reproduce. The steady state is reached when the effective birth rate is balanced by the death rate. In Appendix A we estimate that with the values of parameters used here the total coverage should be approximately 5000, which is in good agreement with the curves of Figure *2a*. Figure *2c* shows the population dynamics of a homogeneous system according to the "global" model. Notice that a declining phase appears as a result of the fact that individuals regulate their reproduction rate based on information from the entire society. If the fittest parts of the society stop reproduction due to local overcrowding then the reproduction rate of all the society drops, although there may be empty space in several locations.

(ii) *Extinction of one of the species*. The second characteristic time is the time needed for the extinction of one of the two species. From the point of view of the population dynamics, we expect the coexistence of both species as they are identical in initial concentration and fitness and know the same number of words. However, when we allow the system to evolve for a long time we observe that usually by the end of the run only one of the two species survives. This is shown in Figure *3*. Figure *3* gives the 2-species "survival" probability i.e. the proportion of runs (out of 150 runs) for which both species still survive at time *t*. From an initial value of 1.00 this probability drops exponentially. The rate is 3.7 x $10^{-6}$ $(MCS)^{-1}$, using the typical 100×100 lattice. Note that when we used a smaller 50 × 50 lattice, then the rate was 1.95 x $10^{-5}$ $(MCS)^{-1}$ (figure not shown). Concerning the survival probability curve for the "global" model the exponential decline has a slope of 3.3 x $10^{-5}$ $(MCS)^{-1}$. The measured time-constants are 50200 MCS for the "local" model, and 10000 MCS for the "global" model. This behavior is reminiscent of the poisoning effects noticed in the A+B→0 reactions associated with noise-induced kinetic phase transitions [11]. Another way of viewing this is in terms of demographic stochasticity [12]. Because the two species are identical in every way, given a constant total population, the only interaction is through the fact that any increase of $N_A$ must be matched by a decrease of $N_B$. This is the "gamblers ruin" scenario, which constitutes a random walk to extinction [13]. Appendix B develops a model using this approach. Provided the number of states is large, then the probability associated with coexistence (states 1 to *K*-1) falls exponentially. For the "local" model the time constant is found to have a mean value of 50000 MCS, which is in excellent agreement with the time-constant 50200 MCS that is observed in Figure *3*. According to the model of Appendix B the above times increase when we increase the size of the simulated system but are less sensitive to the initial species concentration, which is what we observe here. In the present study we are in the situation where the two species coexist.

(iii) *Establishment of a common language*. We notice that the fitness per individual is constantly increasing with time for both species. This is not done in the quasi-linear way observed in Figure *1b* because now the parameter *C* in Eq. 3 is no longer a constant, as individuals are born and die. This increase in fitness is conferred by communication and is associated with an increase in the number of words. Figure *4* shows a plot of the average number of words of the *A* language that members of the *A* and *B* species know and the average number of words of the *B* language that members of the *A* and *B* species know. As expected, no language prevails over the other and the result is a population of two species that speaks both languages. This is a situation that resembles the linguistic situation in New Guinea, for example, where in an area of 800,000 $km^2$ there are about 1000 different languages [14] (not dialects), and where there are several individuals speaking more than one language. The important time constant here is the time needed for the development of a common language. For a 100x100 lattice with initial concentrations $c_A=c_B=0.15$ and initial fitness $f_A=f_B=30$ units and interactions according to the "local" model, this time is found to have a mean value of 48900 MCS (±12100 MCS). The common vocabulary has a mean value of 19.96 words. The population ends up speaking both languages equally well. Using the "global" model with the same initial condition we have found the mean value for the development of a common language to be 9330 MCS (±2420 MCS). The common vocabulary had in this case a mean size of 19.75 words.

**b. Inhomogeneous System**

In the previous case the system was “homogenous” because the species were identical regarding initial fitness and population density. Figure *5* presents results for the population dynamics (Fig 5a), the average fitness per individual (Fig 5b) and word evolution (Fig 5c) for the same system when the initial fitness of species *A* is 100 units and that of species *B* is 30 units.

(i) *Population Dynamics*. The form of the curves for the number of individuals for species *A* and *B* resemble the homogenous case. The main difference is that the steady value of *B* is lower than that of *A*. The initial fitness advantage given to *A* allows it to grow more rapidly than the *B* species and to take a greater share of the arena.

(ii) *Extinction of one species*. The drift to extinction, observed in the homogenous case, is also observed here. This is expected to be greater (favoring persistence of *A*) since the system rapidly arrives at a state closer to the boundary.

(iii) *Fitness and Language Evolution*. We notice that the average fitness per individual rapidly becomes equal for both species and remains so for a long time. Again, this fitness pattern reflects the evolution of vocabulary, shown in Figure *5c*. In this time regime language *A* seems to dominate as the average number of *A*-language words known to an individual is higher than those of the *B* language. However, the *B* language survives. If we monitor the system with the initial conditions presented in Figure *5* for a much longer period, then a common language is formed (in this case this mean time is 51500 MCS), where the vocabulary has an average size of 19.80 words, i.e. the population will finally become bilingual.

An interesting case appears when the initial concentration of the *B* species is much less than that of the *A* species and also the initial fitness of *B*'s is much less than that of the *A*'s. In Figure *6* we present plots of the number of individuals vs. time (Fig 6a), the average fitness vs. time (Fig 6b) and the number of words vs. time (Fig 6c) for such a "highly inhomogeneous" system with initial concentration $c_A$=0.15 $c_B$=0.05 and initial fitness $f_A$=2000 and $f_B$=30, using the "local" model to describe interactions. The number of *B* individuals is rapidly decreasing. The much higher initial fitness of the *A's* allows them to fill more sites faster surrounding the *B's* and not leaving any room for the B reproduction. Although the fitness advantage is once again balanced rather quickly as shown in Figure *6b*, it is decisive in creating a spatial arrangement that practically eliminates the chances of B reproduction. If we study the same system using the “global” model the average number of B individuals drops very quickly to zero (figure not shown). We observe that although the B population is rapidly becoming very small or even extinct in several cases, part of their vocabulary survives. Thus, for the “highly inhomogeneous” system presented in Figure *6c*, the situation is rather different from that presented in previous figures. Notice that after 400 MCS an *A* individual knows on average 1.5 words of the *B* language (besides his *A* language words) and after 5000 MCS he knows 2.5 words of the *B* language on average. The population will not arrive at a final bilingual state as in the previous case but at a state where only few words of the B language survive.

We may interpret the result as follows: Initially there were two separate languages referring to the same set of objects. For example, individuals of the *A* species might use word number *1* of the lexicon array to identify a certain object while the *B*’s used word number *11* to identify the same object. At the end individuals of the *A* species know two words in order to identify the same object. A richer language containing *synonyms* has emerged from the two distinct languages. The presence of synonyms in a language is not easily justified from an evolutionary point of view as it increases complexity without actually improving communication. The above mechanism shows a way to arrive at a language containing *synonyms* starting from two separate languages, even if one of the two languages is spoken by a small fraction of the population and even if this fraction in most cases disappears rather quickly!

**c. Segregation**

We have mentioned that segregation plays an important role in the behavior of the homogeneous as well as the inhomogeneous system described above. In Figure *7* we present two snapshots of a homogeneous system with variable number of individuals (as described at the section *b* of the Results) at time *t= 1* MCS and at time *t=10000* MCS. The initial concentration of the species is $c_A$ =0.15 and $c_B$ =0.15. The initial fitness of the *A* species is $f_A$ *=30* units and the initial fitness of the *B* species $f_B$*=30* units. Notice the segregation in the second snapshot that is obvious as most of the *B* individuals are surrounded by *A* individuals.

In order to quantify the concept of segregation we use the idea of a segregation coefficient $Q_{AB}$ introduced by Kopelman [15] and Berry [16]. It is based on a quotient of effective pair-correlation functions

$$Q_{AB} = \frac{N_{AA} + N_{BB}}{N_{AB}} \left(\frac{2\, c_A\, c_B}{c_A^2 + c_B^2}\right) \qquad (4)$$

where $c_A$ and $c_B$ is the density of A and B species respectively, $N_{\mathrm{AA}}$ is the number of *A–A* pairs, $N_{\mathrm{BB}}$ is the number of *B–B* pairs, and $N_{\mathrm{AB}}$ is the number of *A–B* pairs. A pair is defined here by two nearest neighbor sites. $N_{\mathrm{AB}}$ for instance, is thus the number of site pairs of which one site is occupied by *A* and the other by *B*. A random distribution of the individuals over the lattice yields $Q_{AB}$= 1, whereas under *A–B* segregation, most of the *A–B* pairs are located on the interfaces between *A*-rich and *B*-rich domains, so that $Q_{AB}$> 1. Thus, the physical meaning of the segregation coefficient is to provide a practical measure of the departure from randomness for the spatial configuration of a system comprised of two ingredients.

In Figure *8* we present a plot of the segregation coefficient versus time keeping the initial concentration of *A* and *B*'s equal to *0.15* and changing the ratio between the initial fitness of *A* and *B*. The plot is generated as follows: At each time step we average over 500 realizations of the system to derive the quantities $N_{AA}$, $N_{BB}$, $N_{AB,}$ $c_A$ *and* $c_B$, and use Eq. *4* to calculate the segregation coefficient.

We notice that for the values of parameters used here there is always segregation and that the segregation coefficient increases with time. It may seem interesting that the value of the segregation coefficient is greater when the initial concentrations and fitness are the same. However, this is a result of the particular definition of the segregation coefficient and its sensitivity on the certain types of spatial arrangement of the species. It is a measure for deciding that the spatial arrangement of a system of two ingredients differs from complete randomness, but it is not actually allowed to compare two different systems and conclude that one is more segregated than the other. What characterizes the spatial arrangement of the "inhomogeneous" cases is the presence of several single B individuals surrounded by A's. Each such arrangement increases the denominator $N_{AB}$ by four units and thus the segregation coefficient is reduced, compared to the homogeneous case where this type of arrangement is much rarer.

## CONCLUSIONS

We have investigated the population dynamics and language evolution of a system of two species speaking two different languages. We considered the population dynamics and the explicit spatial configuration on a two-dimensional spatial lattice. We assumed that species could exchange words ("learn each others language") and bestowed fitness in terms of the number of common words between interacting individuals. We considered cases with and without an initial genetic advantage.

The spatial and time dynamics of the system have rather complex behavior, characterized mainly by the appearance of segregation. The segregation coefficient $Q_{AB}$ was calculated in several cases and found that for a large domain of initial fitness ratios $Q_{AB}$ increases to a final value in the range between 2 and 2.5 at the end of 6000 MCS.

Often, the system will arrive at a fairly stable state, where two species coexist but even when the two species start off as equal, one of them may be lost due to demographic stochasticity. On the other hand, when one of the species consists of rather few individuals with small initial fitness then it, almost always, quickly becomes extinct. The exchange, however, may leave a residue in the language of a dominant species even after the extinction of the other species. A part of its vocabulary has made the final state of the system somewhat "richer" in language, with the acquisition of new *synonyms.*

**Acknowledgements**

This work was partially supported by the Greek Ministry of Education via HRAKLEITOS and PYTHAGORAS Projects.

## Appendix A

*How much of the Arena is Occupied?*

We can consider the spatial grid on which individuals move as a grid of islands that can be colonized or lose their occupants. This problem has been extensively studied by metapopulation theory. According to this theory [17], a collection of identical islands, or lattice sites, can be modeled by binary states: either they are empty or colonized. Individuals move from one site to the next and start colonies if the sites are empty. Using this theory and ignoring the role of language, we can model the dynamics of $x$, the proportion of the lattice occupied:

$$\frac{dx}{dt} = \beta x \cdot (1-x) - \mu x = [\beta \cdot (1-x) - \mu]x \qquad \text{(A1)}$$

The first term on the RHS is the number of new individuals and the second is the number that die. Note that while the number of new individuals produced is $\beta x$ (with $\beta$ the birthrate), the probability of an individual landing on an empty cell is only $1-x$.

At steady-state, the proportion is static, $x=x^*$, for which the equation can be solved. Unless the population is extinct, the proportion of cells occupied is:

$$x^* = 1 - \mu / \beta \qquad \text{(A2)}$$

Thus, if $\beta = 2\mu = 0.1$ MCS, about ***half*** of the arena is occupied. Although this model assumes global mixing, it gives us a good idea of the kind of effect of grid-saturation upon our system.

## Appendix B

*Time to Extinction*

Suppose the number of individuals (occupied cells) in the arena has reached a steady-state, $n_B+n_A=K$, the numbers of either population can range between 0 and $K$. Assuming that there is no essential difference between the two species (neither is a better competitor) $n_A$ , the number of A, may either to increase or decrease with equal probability. Thus, the system performs a random walk. The two boundaries ($n_A$=0 and $n_B$=0) are absorbing. Once either boundary is reached only one species remains and no further change in proportion is possible. This is an instance of the "gambler's ruin problem" [13, p49]. This problem can be solved in a variety of ways. The simplest uses the methods birth-death processes (see, for example, [12]).

Suppose we consider the problem as a birth-death process for which the probability $p_k(t)$ associated with the *k*th state at time *t* is given by the following probability rate-equation:

$$\frac{dp_k}{dt} = -2bp_k + bp_{k-1} + bp_{k+1} \qquad \text{(B1)}$$

Here *b* is the rate at which changes of state occur, the "turnover rate". In a steady state environment, state changes can only happen when one individual dies and another takes its place; changes may be up or down with equal probability. Eq. B1 holds for states *k*=2 to *k*=*K*-2. The fact that the end-states are absorbing means that probability is lost from either end at a rate given by:

$$\frac{dp_0}{dt} = +bp_1 \quad , \quad \frac{dp_K}{dt} = +bp_{K-1} \qquad \text{(B2)}$$

The equations for $p_1(t)$ and $p_{K-1}(t)$ also differ from (B1) because they receive no input from the adjacent absorbing states

$$\frac{dp_1}{dt} = -2bp_1 + bp_2 \ , \quad \frac{dp_{K-1}}{dt} = -2bp_{K-1} + bp_{K-2} \qquad \text{(B3)}$$

This system of equations (B1-B3) can be formally solved, subject to the initial condition $p_{K/2}(0)=1$ (i.e. that the system starts with equal proportions of both species). However a simpler way is to use the "quasi-stationary" approximation [18,19] approach. This assumes is that while the probabilities do change in time, the essential shape of the distribution does not. This approach works provided the "leakage" rate into the absorbing class is small; the distribution of probabilities (the "quasi-stationary distribution") may be calculated as if the absorbing class didn't exist. This implies that the starting state (initial proportions of species) does not play a major role in the outcome. Thus the absorbing boundaries can be replaced by reflecting ones and we then solve for the constant vector $\{\pi_1, \pi_2, \ldots, \pi_{K-1}\}$. This means that we only need to solve the equations

$$\begin{aligned} &\pi_{k+1} - 2\pi_k + \pi_{k-1} = 0 \quad \ldots(a) \\ &\pi_{K-1} - \pi_{K-2} = 0 \quad \ldots(b) \\ &\pi_1 - \pi_2 = 0 \quad \ldots(c) \end{aligned} \tag{B4}$$

The first of these is a 2nd order linear degenerate difference equation, with a solution

$$\pi_k = [C_1 + kC_2](+1)^k \tag{B5}$$

By symmetry it is obvious that $C_2$=0 so that $\pi_k = C_1$. The B4(b-c) also require equality of $\pi_1$ and $\pi_{K-1}$ with $\pi_k$. Thus all non-absorbing states are equal:

$$\pi_1 = \pi_2 = \ldots = \pi_{K-2} = \pi_{K-1} = C_1 = [K-1]^{-1} \tag{B6}$$

In order to solve (B2), we note that

$$p_k = [1 - p_0 - p_K]\pi_k = (1 - 2p_0)/(K-1) \qquad \text{(B7)}$$

So that B2 (a) becomes

$$\frac{dp_0}{dt} = \frac{b}{K-1} - \frac{2b}{K-1} p_0 \qquad \text{(B8)}$$

This is a first order linear differential equation whose solution is exponential. Using Eq. A2 and the fact that for our system $b = \mu$ (a change may occur if and when an organism dies, leaving an empty space), we find that (with $\beta = 2\mu = 0.1$ MCS) then the associated time constant is:

$$\tau_E = \frac{K-1}{2b} = \frac{(1-\mu/\beta)N - 1}{2\mu} \cong 50\times 10^3 \quad \text{MCS} \qquad \text{(B9)}$$

This corresponds to a rate of $2\times 10^{-5}$ $MCS^{-1}$.

**FIGURES**

1. (a) Plot of the average number of words known to an individual vs time. The concentration is $c=0.15$ for both A and B species. (b) Logarithmic plot of the fitness per individual vs. time for several initial concentrations of A and B species (Lines are for visual aid).

2. Plots of the number of individuals versus time (a) and of the average fitness per individual versus time (b), for a system of two species A and B with different languages having the same initial fitness and random initial distribution assuming "local model" interactions. A100x100 lattice was used with initial concentration $c_A=c_B=0.15$ and initial fitness $f_A=f_B=30.$ (d) Population dynamics of the "global" model with initial concentration $c_A=c_B=0.15$ and initial fitness $f_A=f_B=30.$ (Lines are for visual aid).

3. Survival probability curve for a system of two species with local (°) and global (□) interaction. Both species occupy the lattice with the same initial concentration $c_A=c_B=0.15$ and have the same initial fitness $f_A=f_B=30.$ The semilog plot reveals that after an initial time interval where the probability of coexistence is one, it drops exponentially. The slope of the straight line is $\lambda$=-3.7 $10^{-6}$ is for the local and $\lambda=-3.3$ $10^{-5}$ for the global model.

4. Plot of the average number of words versus time for a system of two species A and B with different languages having the same initial fitness and random initial distribution assuming "local model" interactions. A 100x100 lattice was used with initial concentration $c_A=c_B=0.15$ and initial fitness $f_A=f_B=30$. (Lines are for visual aid).

5. Plots of the number of individuals versus time (a) and of the average fitness per individual versus time (b), and of the average number of words versus time (c) for a system of two species A and B with different languages having the different initial fitness and random initial distribution assuming "local model" interactions. A 100x100 lattice was used with initial concentration $c_A=c_B=0.15$ and initial fitness $f_A=100$ and $f_B=30$. (Lines are for visual aid).

6. Plots of the number of individuals versus time (a) and of the average fitness per individual versus time (b), and of the average number of words versus time (c) for a system of two species A and B with different languages having the different initial fitness and random initial distribution assuming "local model" interactions. A *100x100* lattice was used with initial concentration $c_A=0.15$ $c_B=0.05$ and initial fitness $f_A=2000$ and $f_B=30$. (Lines are for visual aid).

7. A snapshot of the system at time $t= 1$ MCS and at time $t=10000$ MCS. Light gray sites are occupied by *A* individuals while black sites by *B* individuals. The initial concentration of the species is $c_A$ =0.15 and $c_B$ =0.15. The initial fitness of the *A* species is $f_A$ $=30$ units and the initial fitness of the *B* species $f_B=30$ units. (Particle size not to scale)

8. Segregation coefficient $Q_{AB}$ versus time for a *2* species system. Initial concentration of *A* and *B*'s is $c_A=c_B=0.15$. The ratio between the initial fitness of A and B is 30/30, 100/30, and 2000/30 respectively.

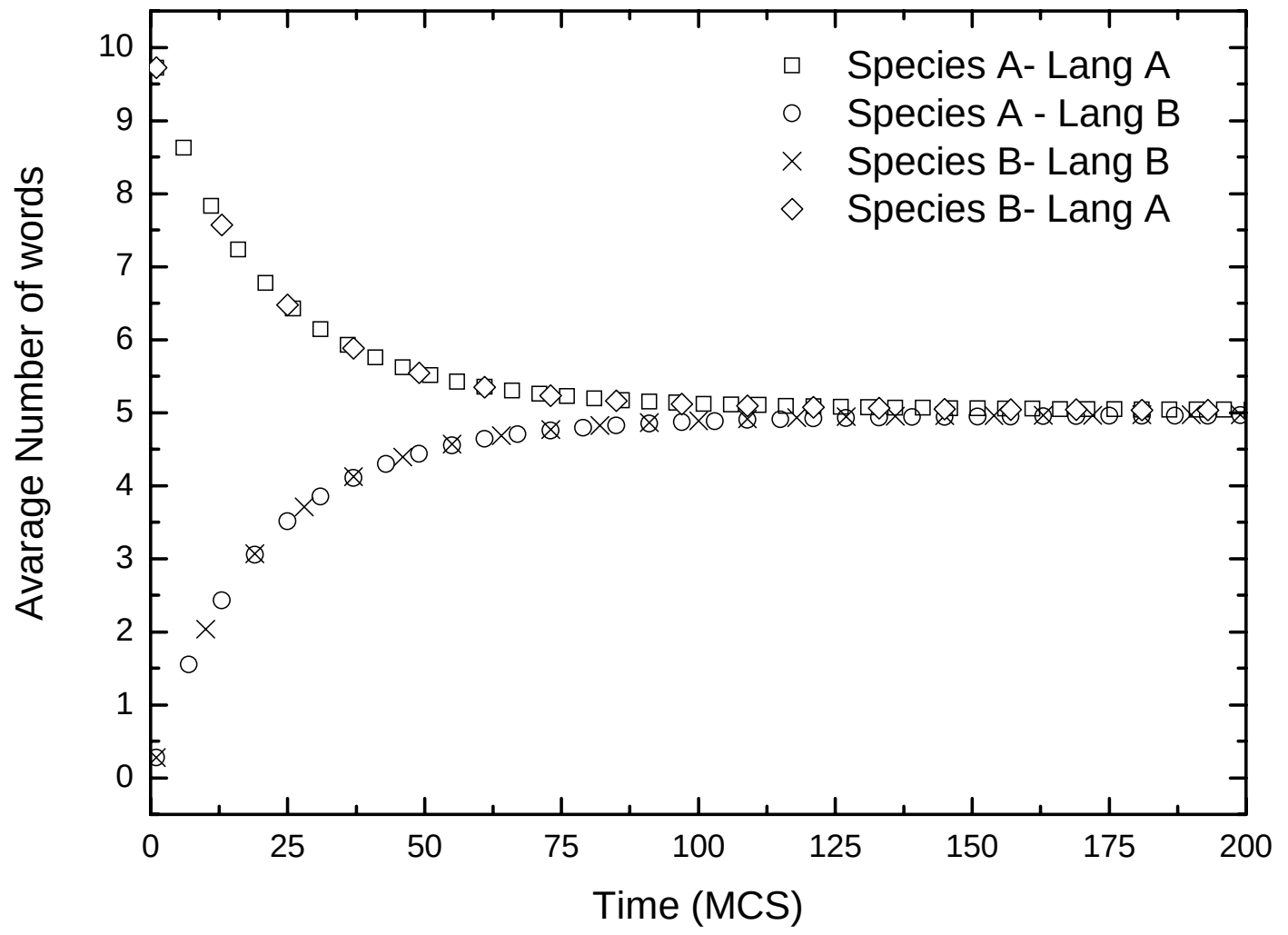

(a)

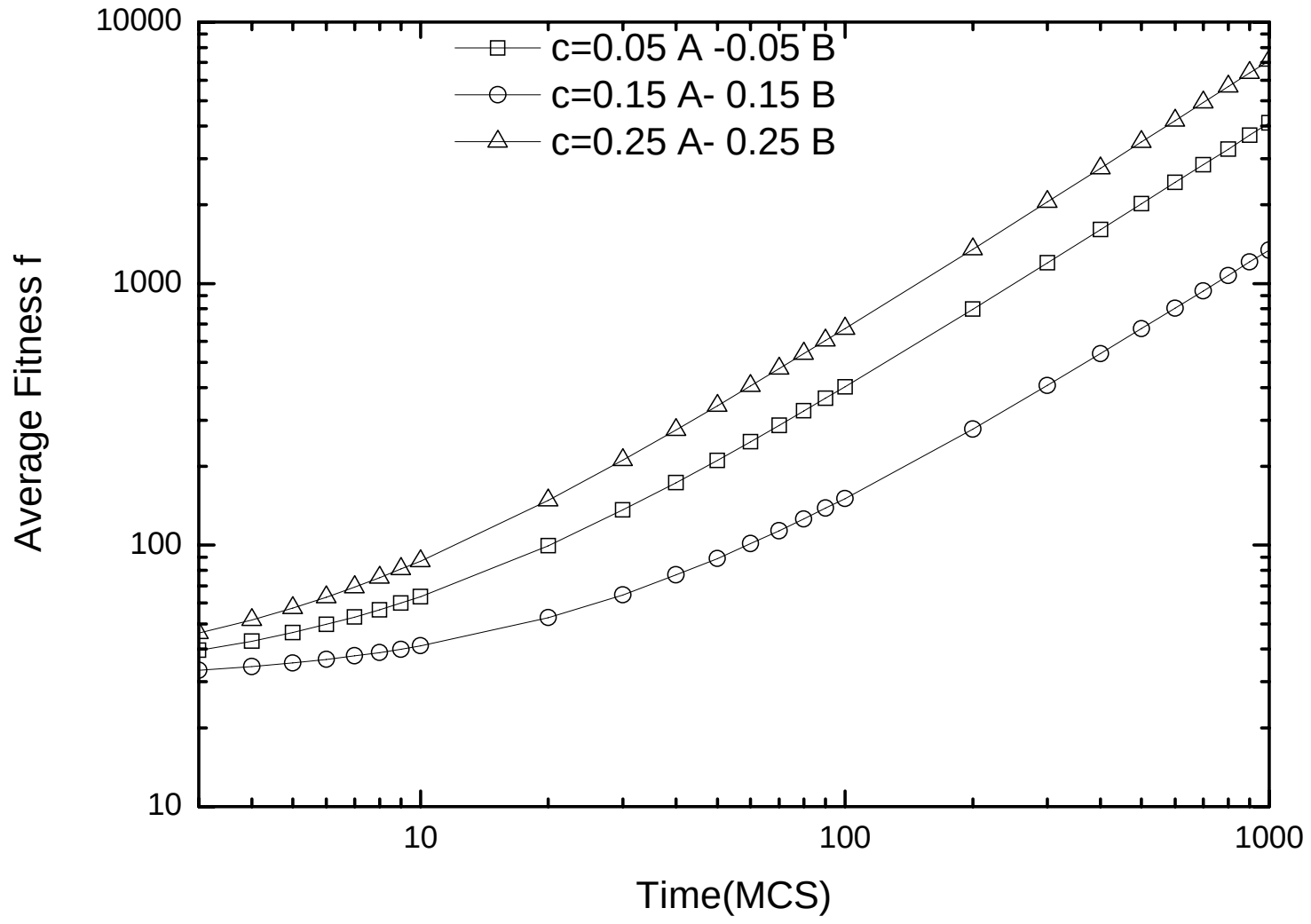
c=0.05 A -0.05 B
c=0.15 A- 0.15 B
c=0.25 A- 0.25 B
Average Fitness f
Time(MCS)
10
100
1000
10000
10
100
1000

(b)

Figure 1

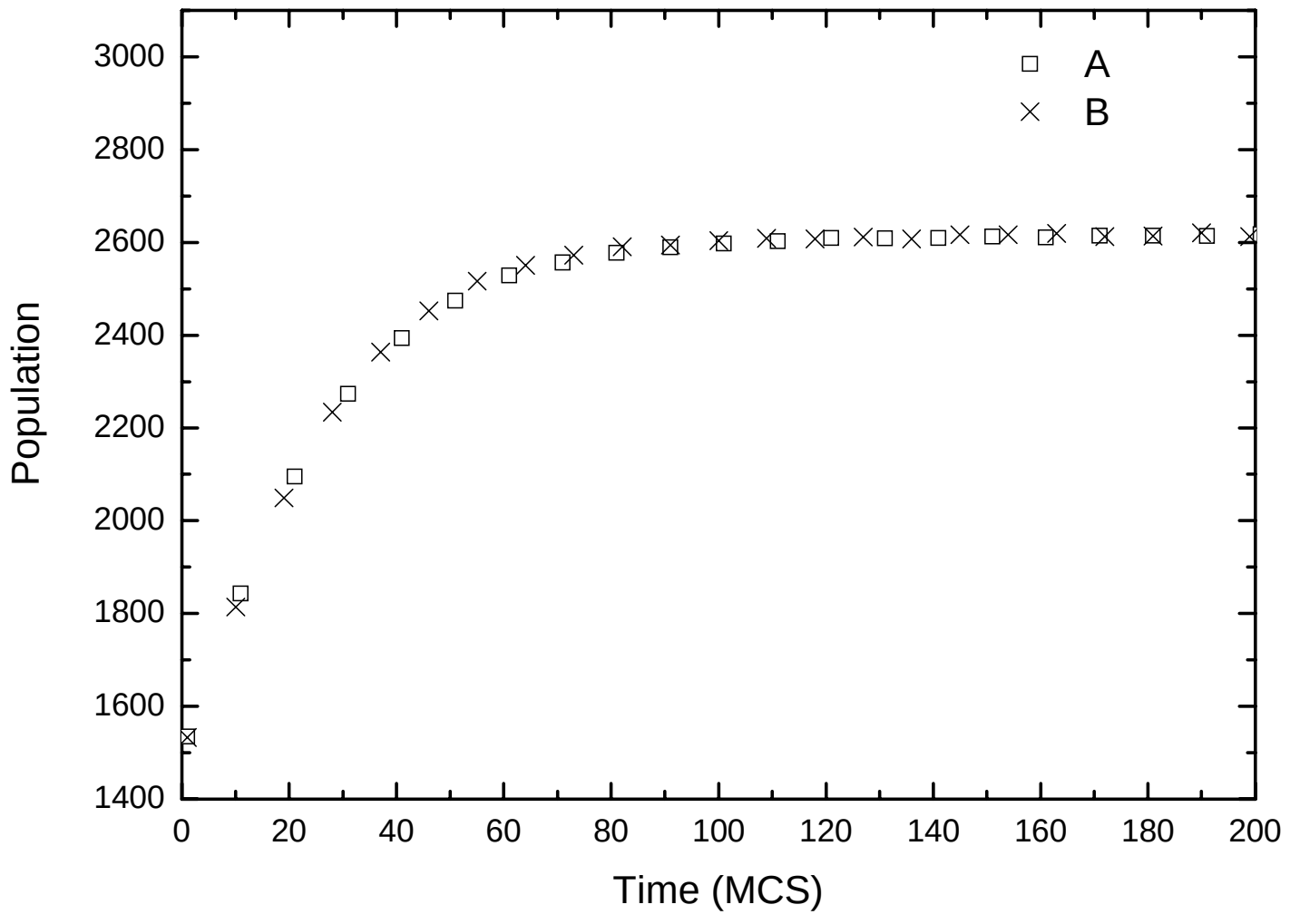

(a)

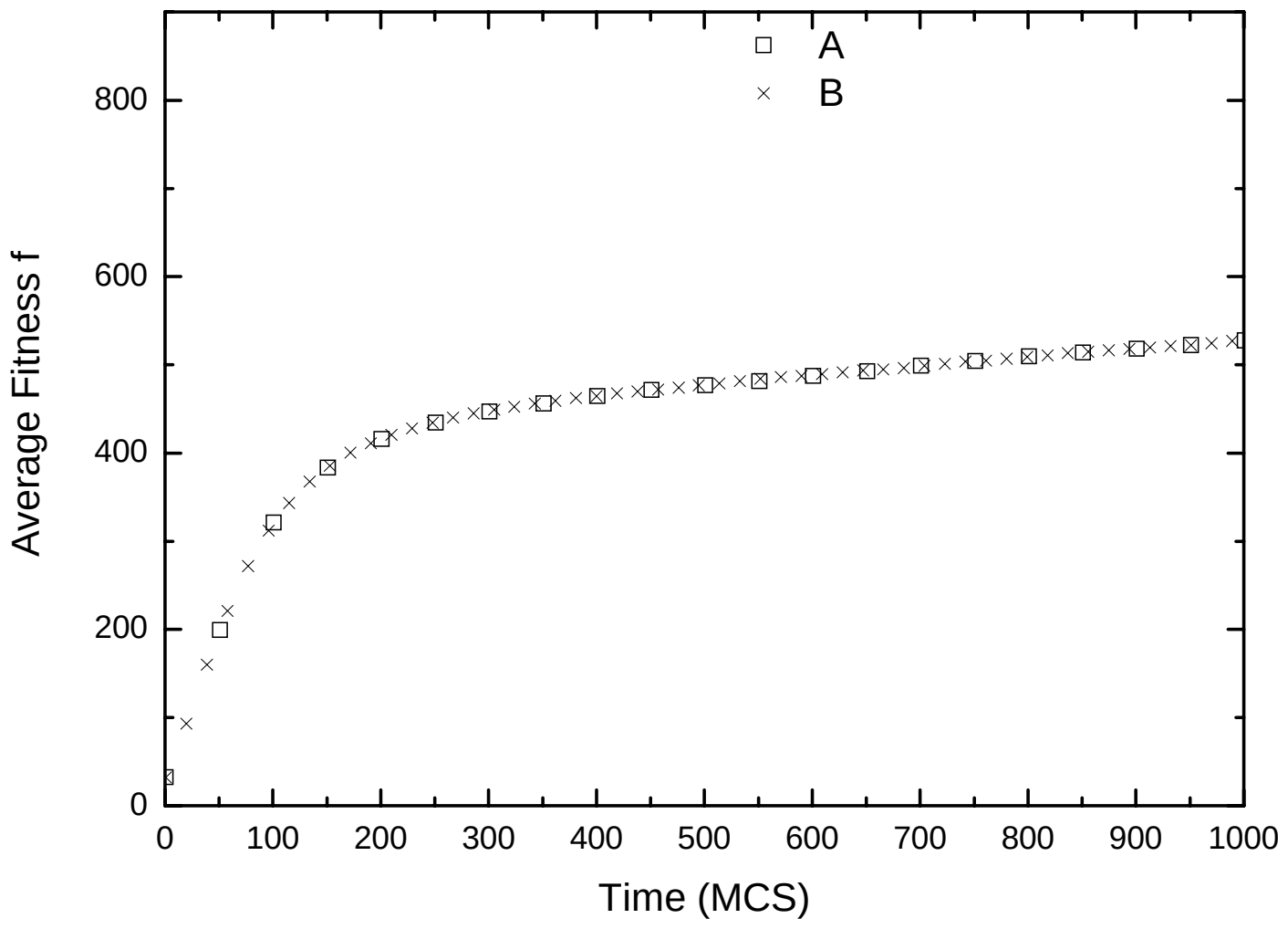

(b)

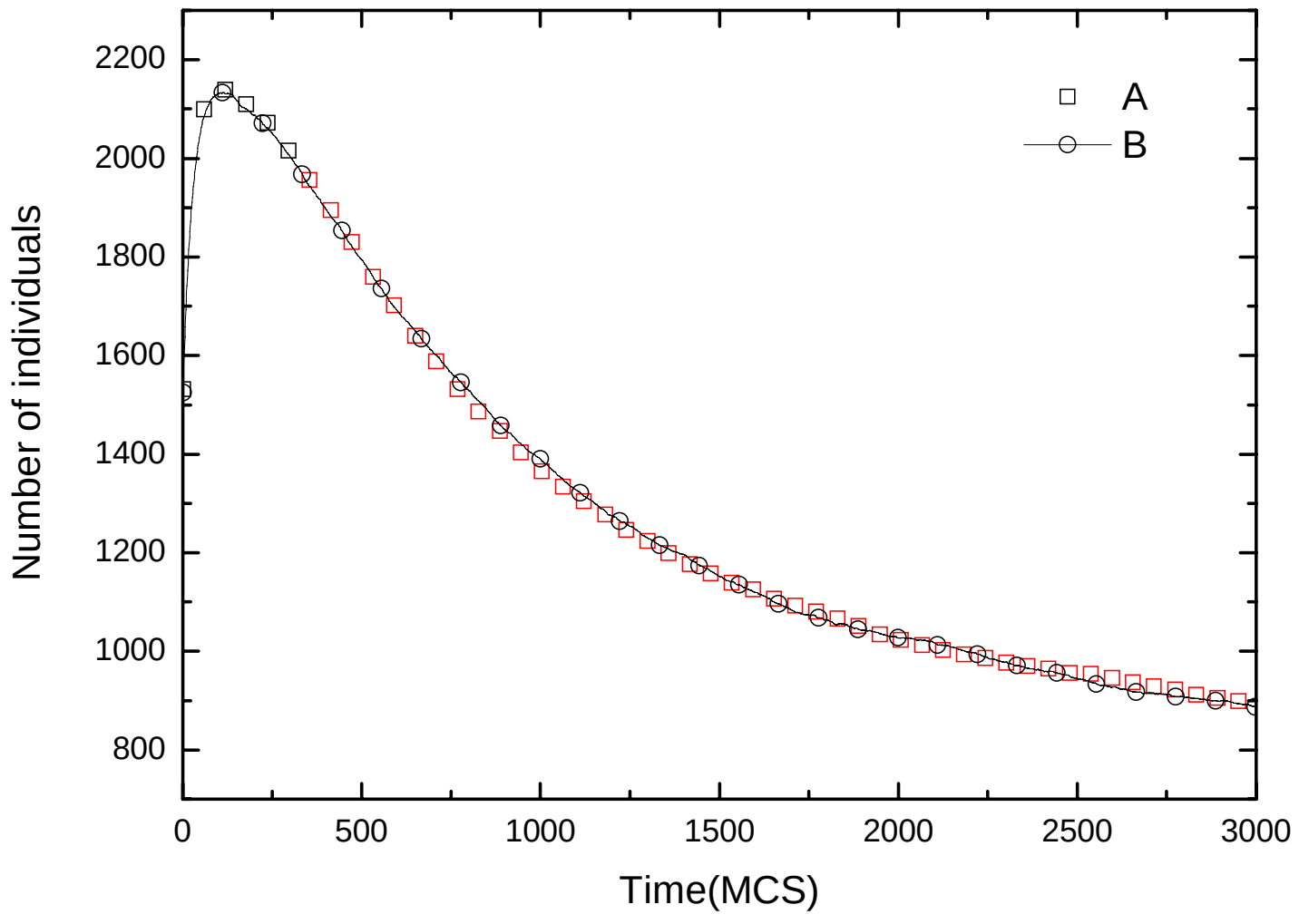

(c)

Figure 2

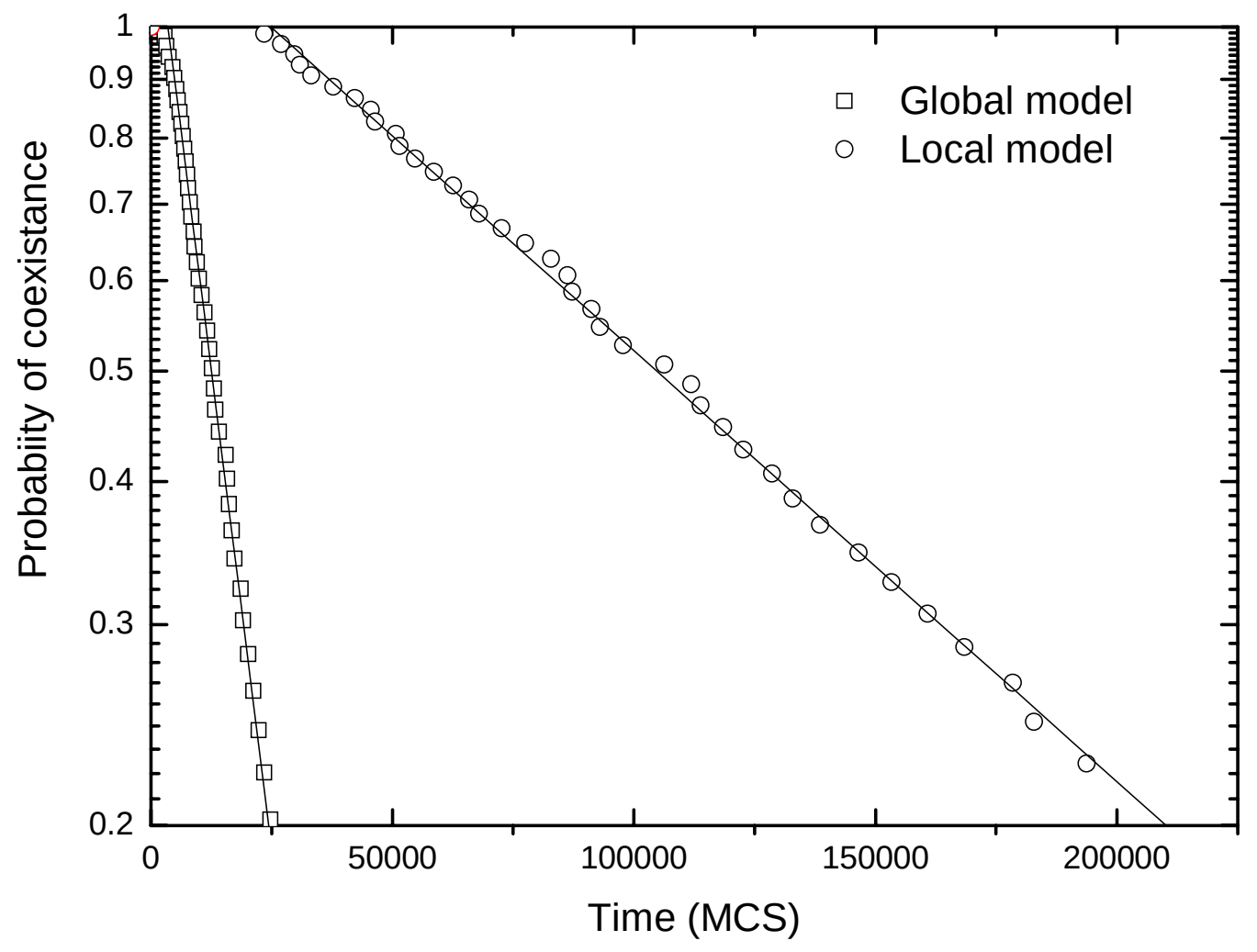

Figure 3

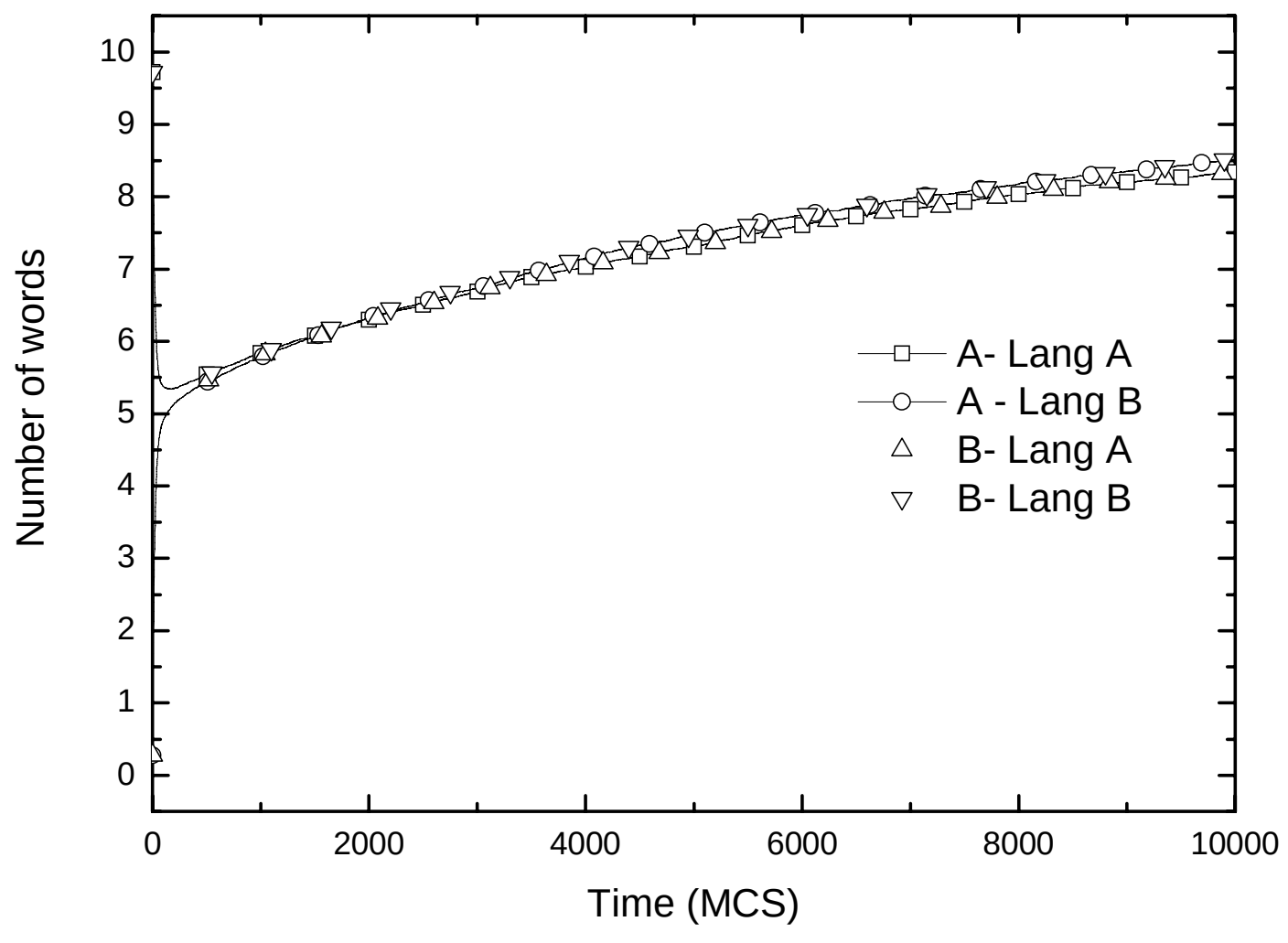

Figure 4

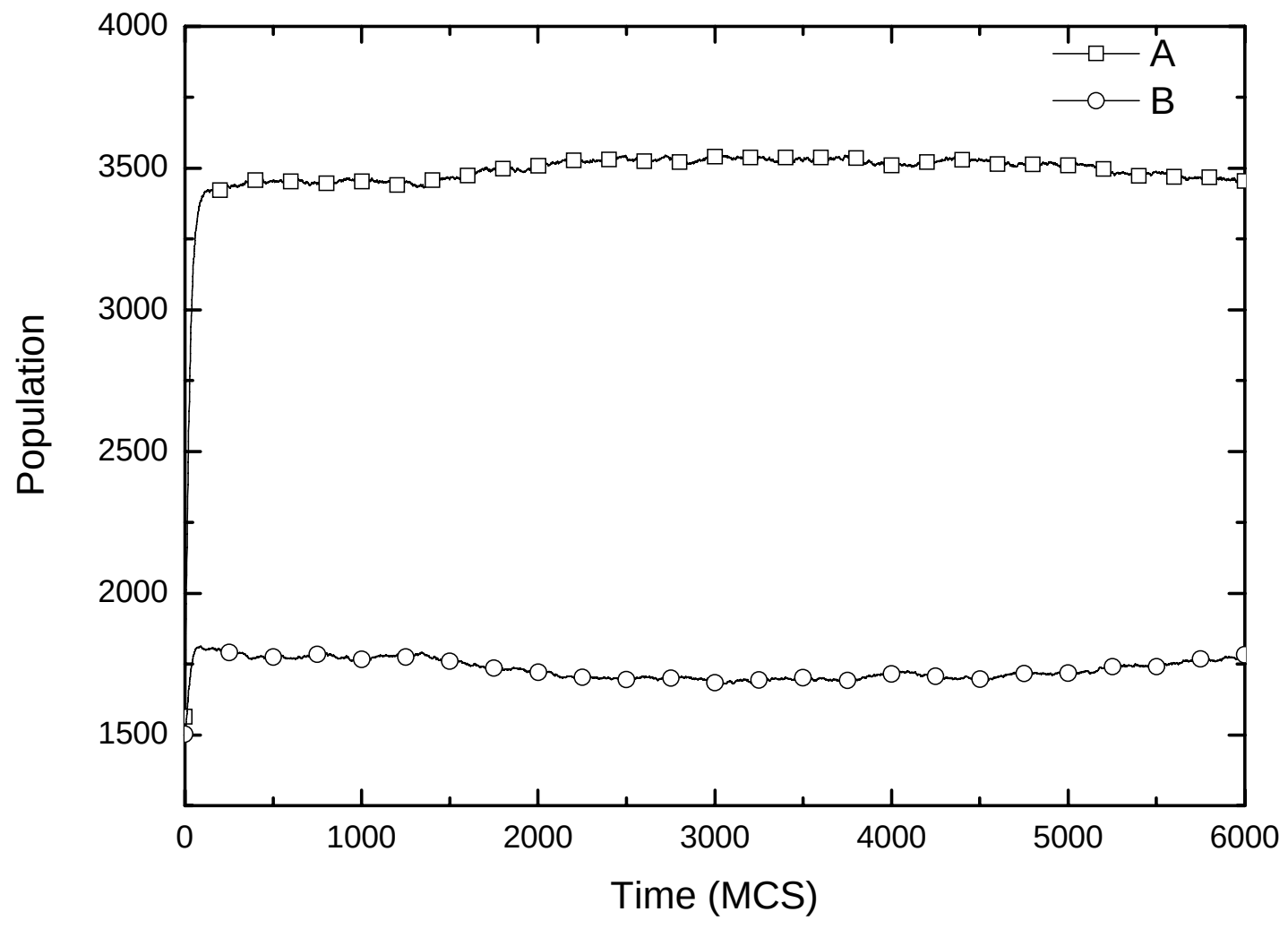

(a)

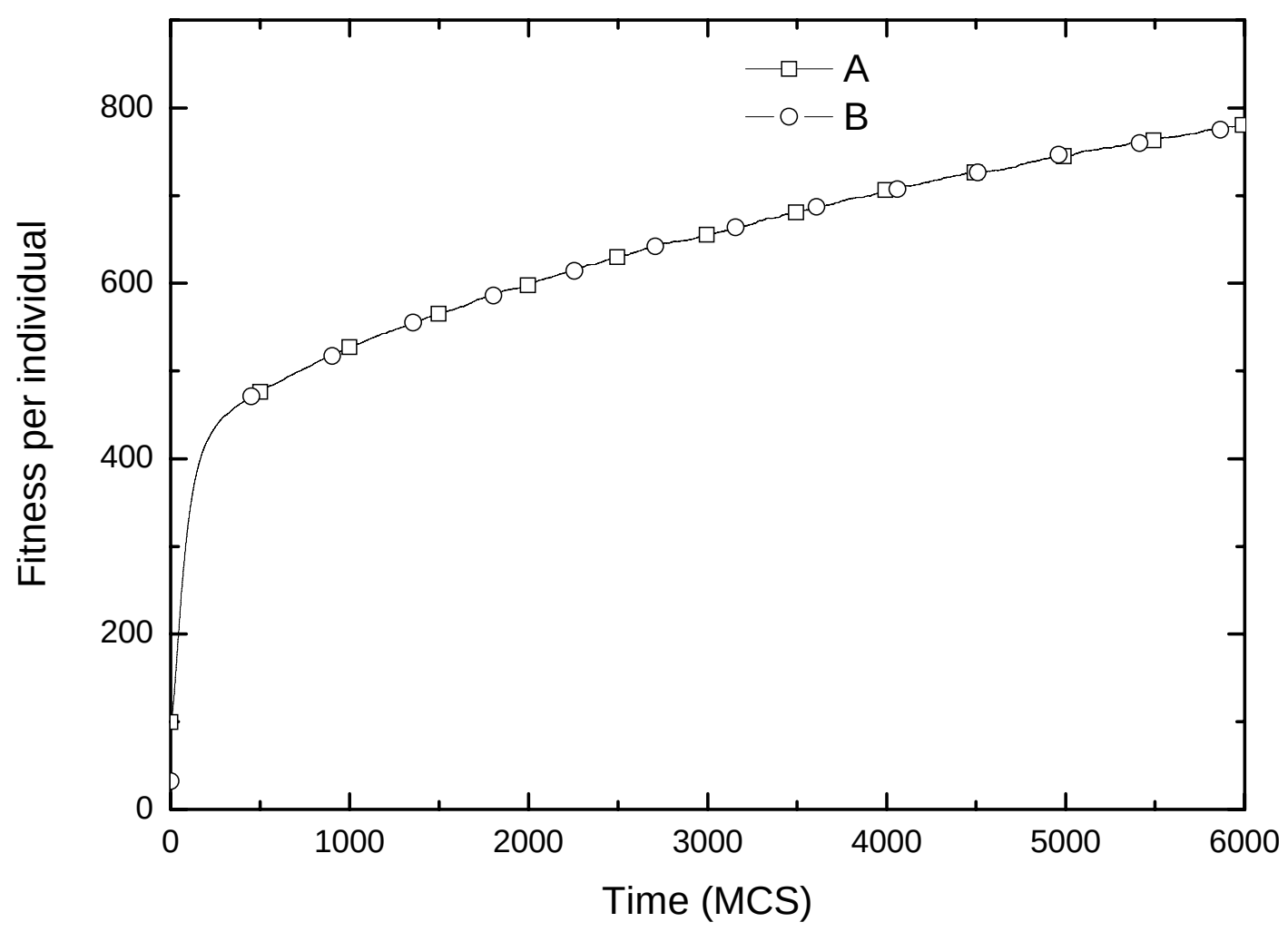

(b)

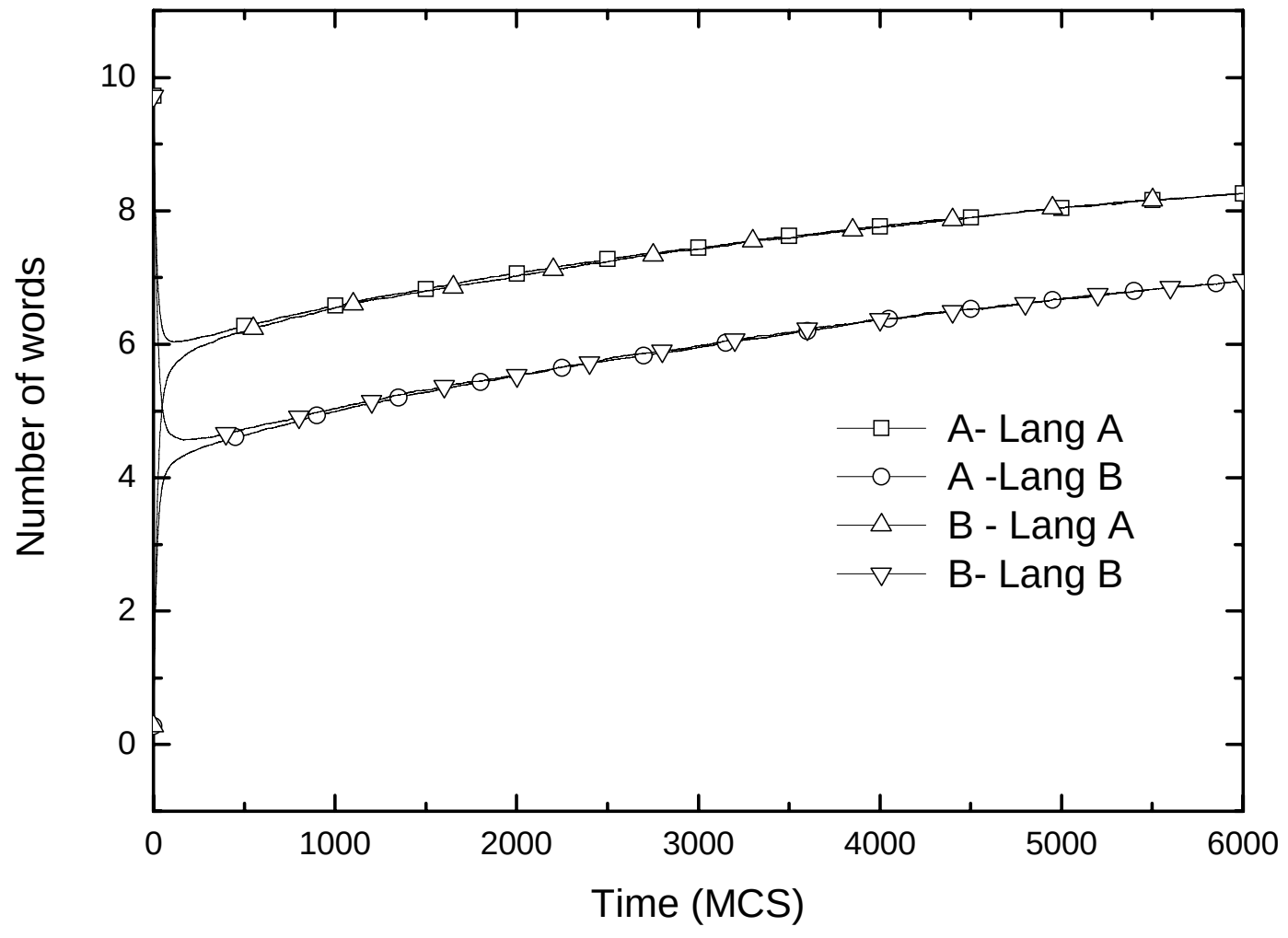

( c )

Figure 5

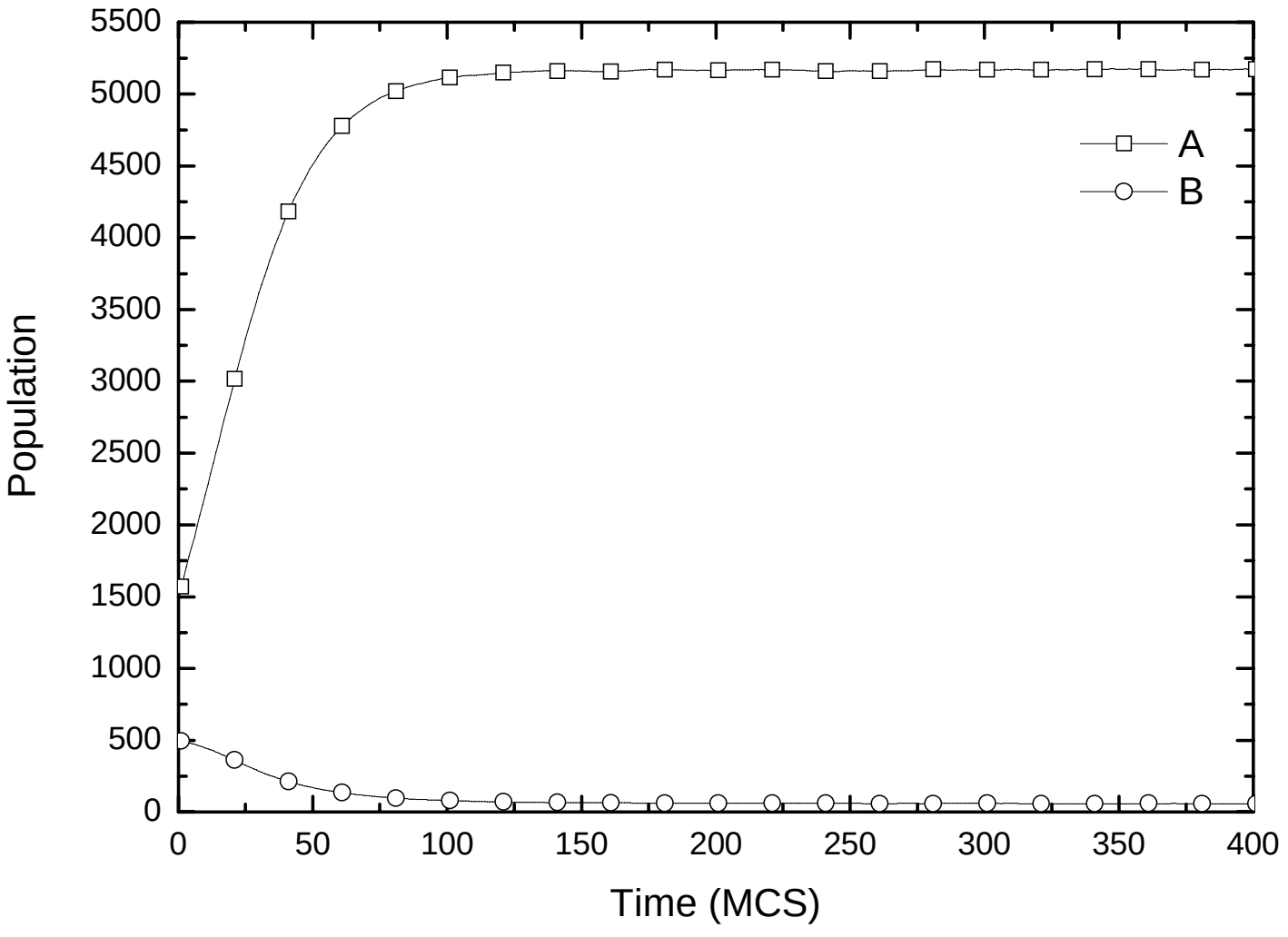

(a)

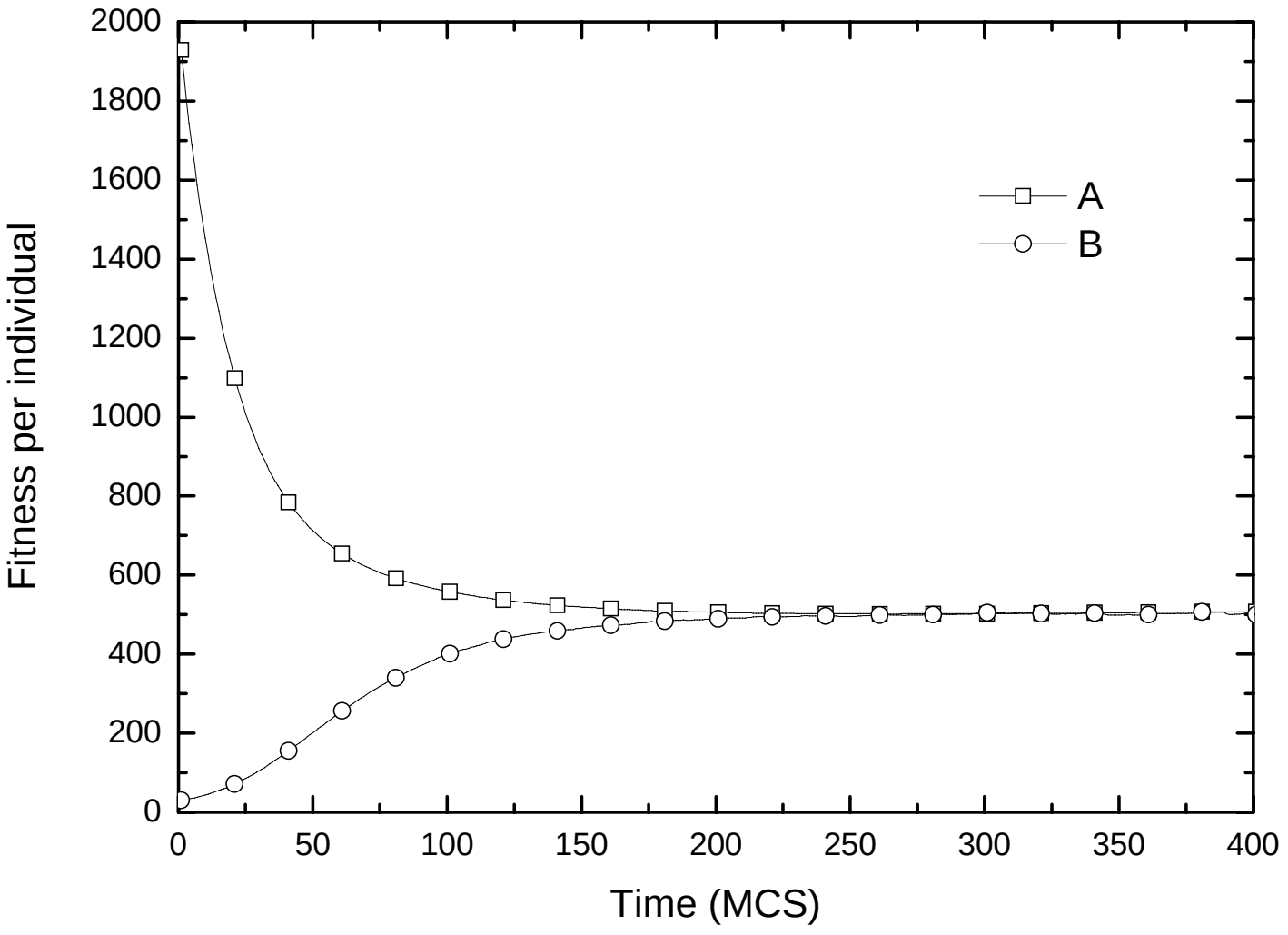

(b)

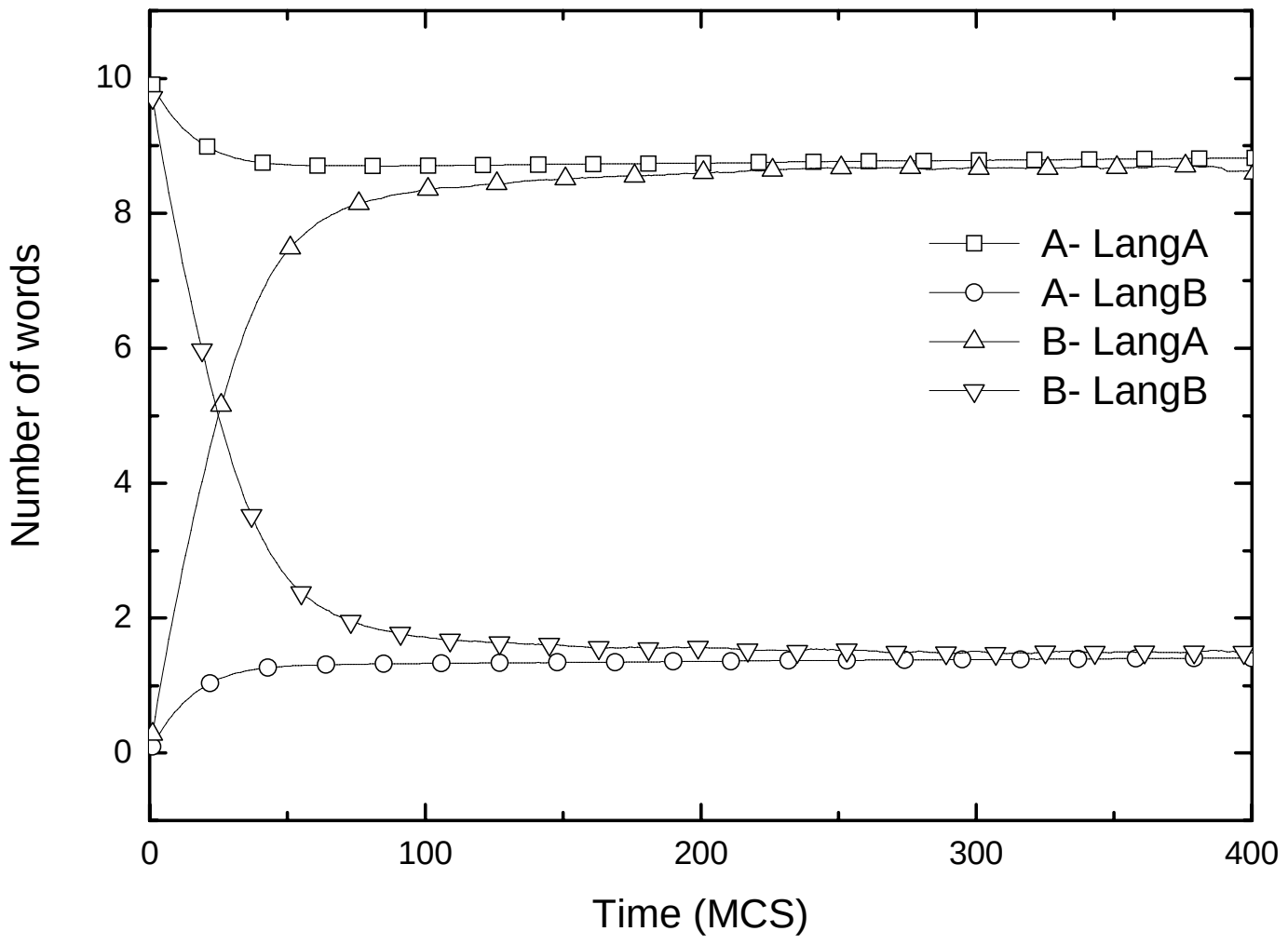

( c )

Figure 6

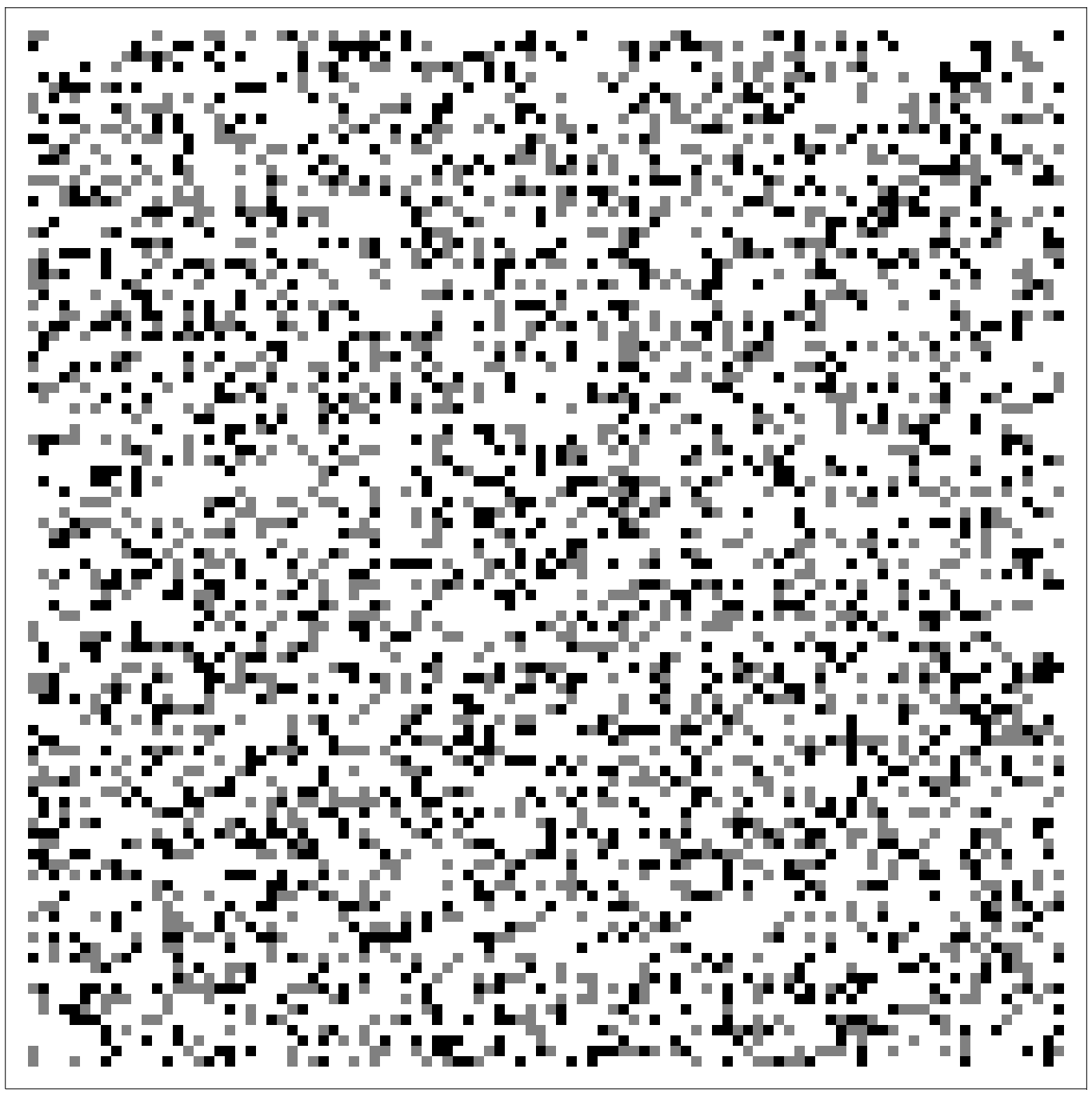

(a)

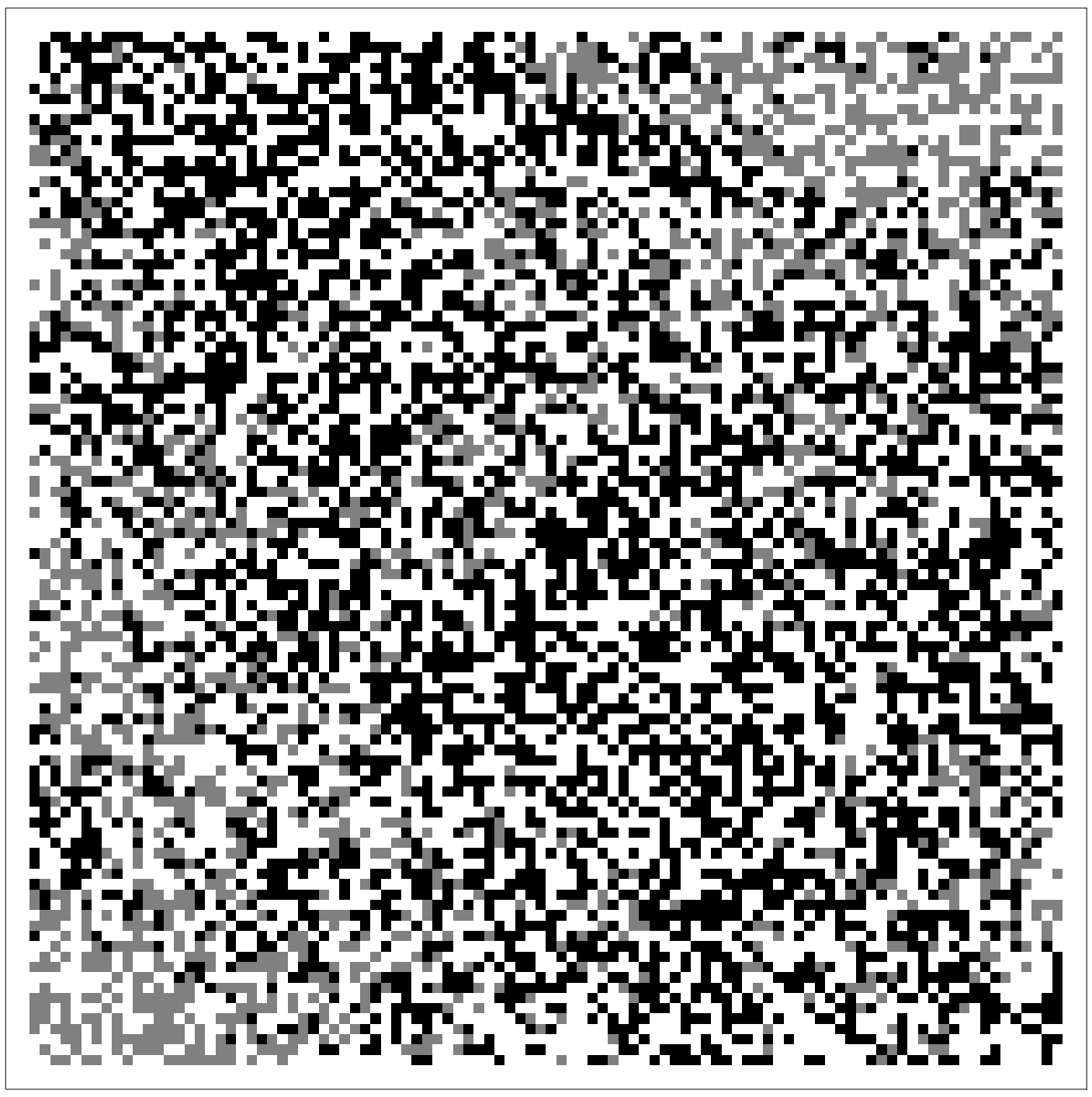

(b)

Figure 7

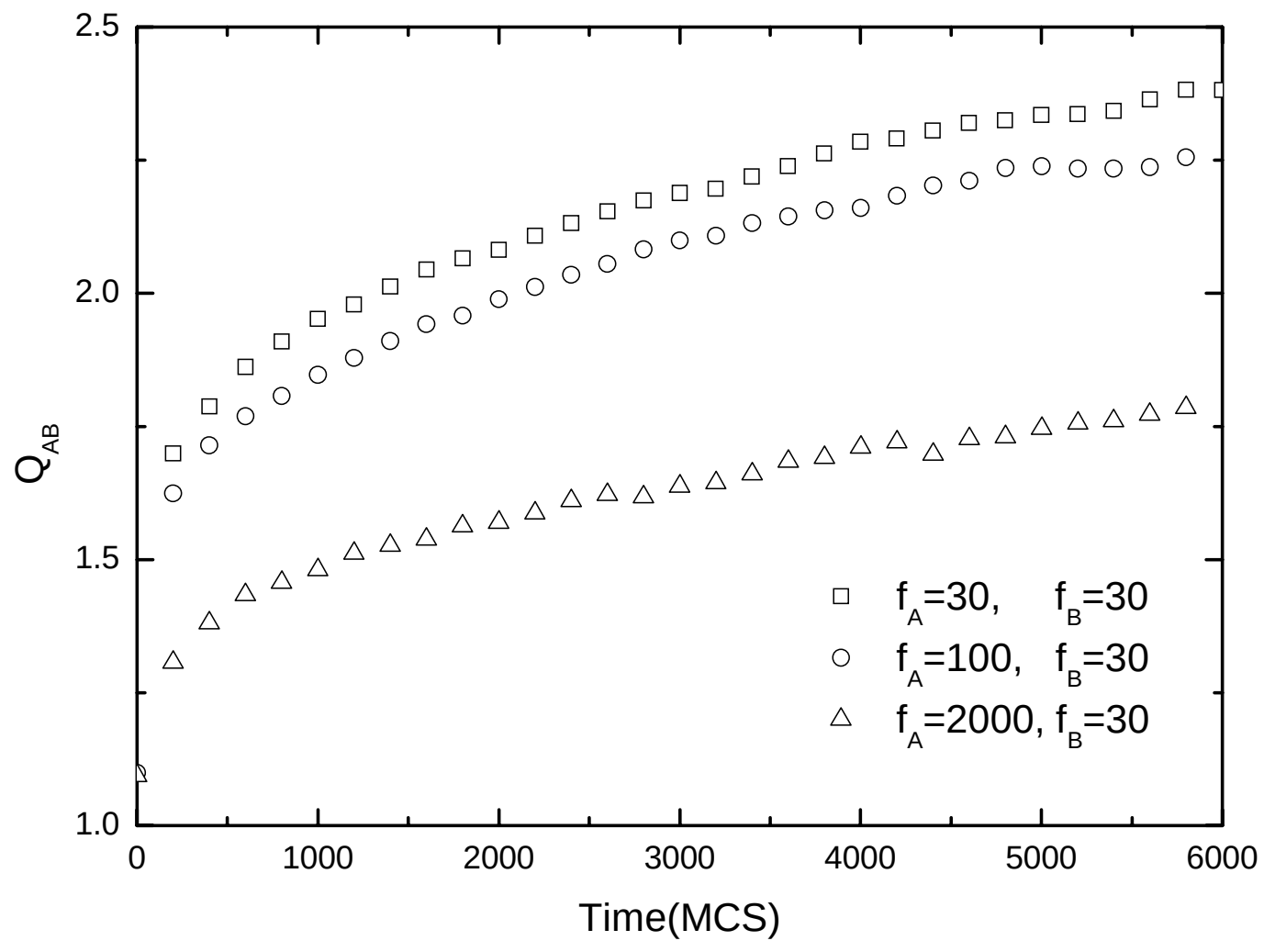

Figure 8